\documentclass{article}

\usepackage{arxiv}
\usepackage{tabularx}
\usepackage[utf8]{inputenc} 
\usepackage[T1]{fontenc}    
\usepackage{hyperref}       
\usepackage{url}            
\usepackage{booktabs}       
\usepackage{amsfonts}       
\usepackage{nicefrac}       
\usepackage{microtype}      
\usepackage{lipsum}		
\usepackage{graphicx}
\usepackage{natbib}
\usepackage{doi}
\usepackage{makecell}
\usepackage{comment}
\usepackage{amsmath}
\usepackage{algpseudocode}
\usepackage{amssymb}
\usepackage{makecell}
\usepackage{algorithm}
\usepackage{enumitem}
\title{WebMCP-Phalanx: Enforcing and Characterizing Trust Boundaries for Browser-Integrated LLM Agents}

\author{ \href{https://orcid.org/0000-0000-0000-0000}{\includegraphics[scale=0.06]{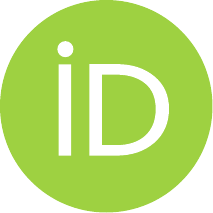}\hspace{1mm}Lin-Fa Lee}\\
	Department of Institute of Artificial Intelligence Innovation\\
	National Yang Ming Chiao Tung University\\
	Hsinchu, Taiwan \\
	\texttt{prologue.ii14@nycu.edu.tw} \\
	\And
	\href{https://orcid.org/0000-0000-0000-0000}{\includegraphics[scale=0.06]{orcid.pdf}\hspace{1mm}Yi-Yu Chang} \\
	Department of Institute of Artificial Intelligence Innovation\\
	National Yang Ming Chiao Tung University\\
	Hsinchu, Taiwan \\
	\texttt{daniel282907@gmail.com} \\
    \And
	\href{https://orcid.org/0000-0000-0000-0000}{\includegraphics[scale=0.06]{orcid.pdf}\hspace{1mm}Chia-Mu Yu} \\
	Institute of Electrical and Computer Engineering, National Yang Ming Chiao Tung University, Hsinchu, Taiwan\\
	Hsinchu, Taiwan \\
	\texttt{chiamuyu@gmail.com} \\
    \And
    \href{https://orcid.org/0000-0000-0000-0000}{\includegraphics[scale=0.06]{orcid.pdf}\hspace{1mm}Kuo-Hui Yeh} \\
	Department of Institute of Artificial Intelligence Innovation\\
	National Yang Ming Chiao Tung University\\
	Hsinchu, Taiwan \\
	\texttt{khyeh@nycu.edu.tw} \\
}

\renewcommand{\shorttitle}{WebMCP Tool Surface Poisoning}

\hypersetup{
pdftitle={A template for the arxiv style},
pdfsubject={q-bio.NC, q-bio.QM},
pdfauthor={David S.~Hippocampus, Elias D.~Striatum},
pdfkeywords={First keyword, Second keyword, More},
}

\begin{document}
\maketitle

The emerging W3C WebMCP standardization proposal enables LLM agents to invoke tools exposed by web pages. In multi-party web environments, however, integrating agent execution into a browser security model centered on the Same-Origin Policy (SOP) raises security concerns that the existing origin-based model does not fully address. In particular, this SOP-centered model provides insufficient provenance and lifecycle guarantees for agent-accessible tools, creating three key risks, i.e. subject-attribution spoofing, uncontrolled tool lifecycles, and semantic prompt injection. To mitigate these risks, we propose WebMCP-Phalanx, a dual-layer agent runtime architecture. The first layer provides a browser-native trust anchor that binds each tool to its registering principal through cryptographically protected capability credentials and propagates provenance labels throughout the tool lifecycle. The second layer separates semantic inspection from privileged tool use. A Quarantine Agent (Q-LLM), which has no authority to invoke tools, examines tool metadata, outputs, and other page-supplied content for prompt-injection attempts. Content that passes this inspection is forwarded to a Privileged Agent (P-LLM) for execution. The Q-LLM’s internal state is not exposed to page scripts, including same-origin scripts, preventing attackers from observing the inspection process and adapting their inputs in response.

Empirical evaluation shows that our browser-native ownership mechanism significantly reduces the success rate of revocation and overwrite attacks from 100\% to 0\%. Furthermore, the dual-agent runtime also blocks all 80 prompt-injection attempts embedded in tool descriptions. For attacks delivered through tool return values, only 2 of 80 attempts remain successful, both caused by tool invocations triggered by malicious tool names before semantic inspection is completed. Across these experiments, the protected runtime preserves task utility, with no statistically significant difference from the no-attack baseline.  Finally, we examine the limits of semantic filtering under a white-box adaptive attacker. The results show that description-based filters can still be bypassed when the attacker places adversarial instructions in tool names and causes the tool to be invoked before inspection. This finding motivates an additional call-timing gate that delays tool invocation until all agent-visible tool metadata has been validated.

\section{Introduction}

The Web is expected to become a major application domain for AI agents. In response to this trend, the W3C is currently developing WebMCP, a draft specification that enables web pages to expose structured tools for invocation by LLM-based agents. This capability enables agents to go beyond simply reading Web pages and to directly execute actions through the tools provided by the corresponding pages. However, this transition raises an important yet largely overlooked concern. As the execution environment of agents shifts toward the more open Web environment, will this introduce additional attack vectors, create new vulnerabilities that make tasks difficult to complete, or even expose limitations in the formulation of current standards?


Motivated by this concern, we conduct a systematic security analysis of the WebMCP specification and validate our findings in a real-world browser environment. WebMCP moves agent execution into the browser's same origin model, where the site's own main program, third party advertising SDKs, and analytics scripts share a single origin without an explicit trust boundary among them. This execution environment reveals four classes of vulnerability that are not merely deployment-level risks but structural weaknesses rooted in the specification itself. The first is subject attribution failure: the tool list carries no origin, so any script can preempt tool names, revoke others' tools, or swap out the tools the agent calls. The second is lifecycle loss of control: tools persist after an SPA logical page transition. The third is execution opacity: a tool can claim to be read only yet exfiltrate data at execution time. The fourth is semantic injection: a structurally legitimate tool embeds manipulative instructions in its description or returned content. These vulnerabilities can be reproduced consistently, are independent of any particular language model, and cannot be adequately addressed through isolated patches. Instead, they require a systematic defense framework based on the WebMCP specification to be thoroughly eradicated. This is precisely what WebMCP-Phalanx aims to provide.


To address these vulnerabilities, we propose WebMCP-Phalanx, a dynamic and secure multi-agent execution environment based on WebMCP standardization proposal. It consists of two complementary defense layers: the browser layer and the multi-agent layer. The browser layer is responsible for establishing the foundation of trust. Tool ownership is provided by the underlying browser, which cannot be forged by any webpage script, ensuring that only subjects holding valid certificates can register or revoke tools. On top of this, several modules separately handle the lifecycle, origin attribution, and execution monitoring, collectively generating a tool trust label. The multi-agent layer, in contrast, performs semantic content inspection on top of the trust labels generated by the browser layer. We employ multiple LLM-based agents to evaluate the trust label from the browser side. Tools labeled as trusted are routed to a privileged agent with tool calling permissions for direct use, whereas suspicious or insufficiently identified tools are first examined by a quarantine agent without such permissions. Before tool descriptions, input schemas, and post execution returned content reach the privileged agent, the quarantine agent first reads and evaluates their semantics, intercepting attacks that are structurally completely valid but hide injection instructions within the semantic layer. By combining these two complementary defense layers, WebMCP-Phalanx addresses both specification-level structural vulnerabilities and semantic injection attacks.




At the browser layer, our defense mechanisms reduce the success rates of tool revocation and replacement attacks from 100\% to 0\% under the evaluated settings, while the proposed capability credentials withstand all tested forgery attempts. However, subject attribution remains only partially addressed since any same origin script obtains a capability of its own once it registers a tool, degenerating the origin dimension. We empirically characterize this limitation and identify the additional attribution information that should be incorporated into the WebMCP specification. The multi-agent layer complements the browser-layer defenses by inspecting the actual semantic content associated with each tool, thereby detecting injection attacks that cannot be identified solely from structural trust labels. We further show that against task fitting tool names, merely deleting malicious descriptions without reading the content fails in proportion to name relevance, whereas letting the agent read the content to judge remains unaffected.


The contributions of this paper are as follows:
\begin{itemize}
\item We systematically identify and characterize the structural security flaws in the WebMCP specification, demonstrating how its current trust and execution model enables several classes of attacks.
\item We propose WebMCP-Phalanx, a secure multi-agent execution environment that combines browser-level trust enforcement with multi-agent semantic inspection to defend against both specification-level vulnerabilities and semantic injection attacks.
\item We reveal that existing multi-LLM defense mechanisms rely on an implicit assumption that the system can correctly distinguish trusted content from untrusted content. We show that this assumption does not hold under the current WebMCP design. Furthermore, we characterize the structural boundary of description layer content inspection under a white box adaptive attacker, across two model families and three classifier prompt designs.
\item Our study is conducted at a critical stage in the development of WebMCP. By identifying these issues before specification ossification, we provide actionable recommendations for strengthening the specification while fundamental design changes remain feasible.

\end{itemize}
\section{Related Work}

\textbf{Security Challenges in WebMCP.} The transition of LLMs to web agents via MCP dismantles traditional single trusted subject security models \cite{servers25mcp, jing2025mcip}. Indirect Prompt Injection (IPI) threats have surged, spanning visual injections \cite{wang2025webinject, cao2025vpi}, conversational manipulations \cite{chen2025topicattack, zou2025poisonedrag, deng2023masterkey, zhong2023poisoning, chang2025chatinject}, and isolated data fragments \cite{wang2025obliinjection, chang2026overcoming}. In WebMCP's multi source environments, attackers can hijack task planning via weaponized tool descriptions \cite{sneh2025tooltweak, wangtrojantools} or pollute subsequent operations \cite{li2025dissonances}. Operating under the browser's flat Same Origin Policy (SOP) without tool ownership verification leaves agents entirely defenseless.

\textbf{Bottlenecks of Model Intrinsic Defenses.} Initial defenses relied on model intrinsic reinforcements, such as structural prompt isolation \cite{chen2025struq}, preference optimization \cite{chen2025secalign}, and runtime masking \cite{he2026attriguard, zhu2025melon}. However, burdening a single model with both complex reasoning and rigorous security filtering exacerbates hallucinations and drops task success rates below 45\% \cite{wang2508mcp}. Relying exclusively on backend semantic validation is fundamentally insufficient to protect contextual integrity at the protocol layer \cite{jing2025mcip}.

\textbf{Multi Agent Privilege Separation \& Our Positioning.} Recognizing these bottlenecks, research shifted toward multi agent privilege separation \cite{brumley2004privtrans}. Frameworks like IsolateGPT \cite{wu2024isolategpt} and ACE \cite{li2025ace} enforce sandbox isolation, while MAS guardrails \cite{kim2025prompt, wang2025g, zhu2025master, de2025open} govern data topologies to prevent privilege escalation. Yet, these sophisticated defenses share a critical flaw: they assume the \textit{a priori} legitimacy of the tool registry. Under the SOP, malicious scripts can trivially spoof tool descriptions; if the registry is compromised natively, tools isolated by the backend are already malicious by design. WebMCP-Phalanx bridges this gap. Intervening at the native browser layer, it issues unforgeable capability handles to establish deterministic ownership. These trust labels are seamlessly propagated to an asymmetric multi agent runtime, coupling native authorization with semantic data routing to systematically eradicate WebMCP's structural vulnerabilities.
\section{Attack Method}

\subsection{Threat Model and Attack Taxonomy}

We consider an LLM agent interacting with a web page through the WebMCP standard. The web page registers agent-accessible tools with the browser, while the agent retrieves the registered tool list, invokes selected tools, and receives their returns via the browser. Crucially, the execution environment operates under the browser's Same Origin Policy (SOP), meaning the site's main program, third party advertising SDKs, and analytics scripts share a single origin. Consequently, these components can access the same WebMCP interfaces despite having different levels of trust. This results in a largely flat execution environment in which the current WebMCP specification provides no explicit trust boundary among scripts operating within the same origin.


\subsection{Attacker Capabilities and Privilege Scope}
In this website, the attacker is a compromised advertising SDK or a tainted analytics script. Operating strictly as a page script sharing the origin of the legitimate site, the adversary possesses the following vector capabilities:
\begin{description}[leftmargin=*]
    \item [Turing Complete Context Control (Arbitrary JavaScript Execution):] As one of the scripts loaded in the page, the attacker can execute arbitrary JavaScript within the page. This includes reading and manipulating the DOM, accessing global objects, and using advanced language level techniques to subvert standard operations.
    \item [SOP Granted Registry Access (Same Origin):] Because the attacker and the legitimate website share the same origin, the adversary is granted access to the identical tool registry. This enables the attacker to autonomously register malicious tools and attempt to revoke existing tools.
    \item [Asynchronous Temporal Exploitation (Unrestricted Timing):] The attacker is not constrained by a deterministic loading sequence and can be loaded before or after the legitimate scripts. This temporal advantage allows the attacker to attempt to preempt, overwrite, or race for tool registration.
\end{description}
We establish a strict cryptographic boundary: the attacker holds a capability credential only for tools it registered itself, and it fundamentally cannot present the credential of another registrant. We specifically scope the adversary to same origin page scripts. A browser extension executes in an isolated world under a distinctly different privilege model, to which the same origin reasoning of this paper does not apply; thus, extensions are out of scope.

The attacker's ultimate goal is to compromise the agent's control flow, causing it to execute harmful actions or forcing the original task to fail. This specifically includes inducing the agent to call malicious tools, exfiltrating the user's sensitive data, or hijacking the agent to perform unauthorized operations.

\subsection{Taxonomy of WebMCP Exploits}
Under the aforementioned adversarial model, the attacks that an attacker can launch are divided into two fundamental categories based on their structural nature. These categories correspond directly to the two complementary lines of defense within the WebMCP-Phalanx framework:
\begin{description}[leftmargin=*]
    \item [Protocol and State Manipulation:]  These exploits target vulnerabilities at the specification level. For example, the adversary can manipulate the state of tools, or exploit incomplete clearance during page transitions to make tools and their residual semantics persist.
    \item [Semantic Poisoning (Content Level Attacks):]  These exploits weaponize the linguistic interface of the LLM by hiding instructions to manipulate the agent within a structurally completely valid tool. Specifically, the adversary conceals these instructions within the tool's description, input schema, or post execution return content.
\end{description}
\section{Proposed Method}

The WebMCP-Phalanx framework is engineered upon a foundational security paradigm: the strict decoupling of origin verification from semantic execution. Within this architecture, the native browser layer acts as a deterministic trust anchor, generating unforgeable trust labels for each tool, while the downstream multi agent runtime dynamically enforces data flow policies based strictly on these labels.
This structural decoupling introduces a critical security property: the autonomous agent is completely relieved of the cognitive burden of deducing a tool's trustworthiness. Because the browser layer cryptographically binds and labels the tool before it reaches the LLM, the agent operates on a provably reliable foundation to isolate and neutralize semantic layer threats.

\begin{figure*}[t]
    \centering
    \includegraphics[width=0.90\textwidth]{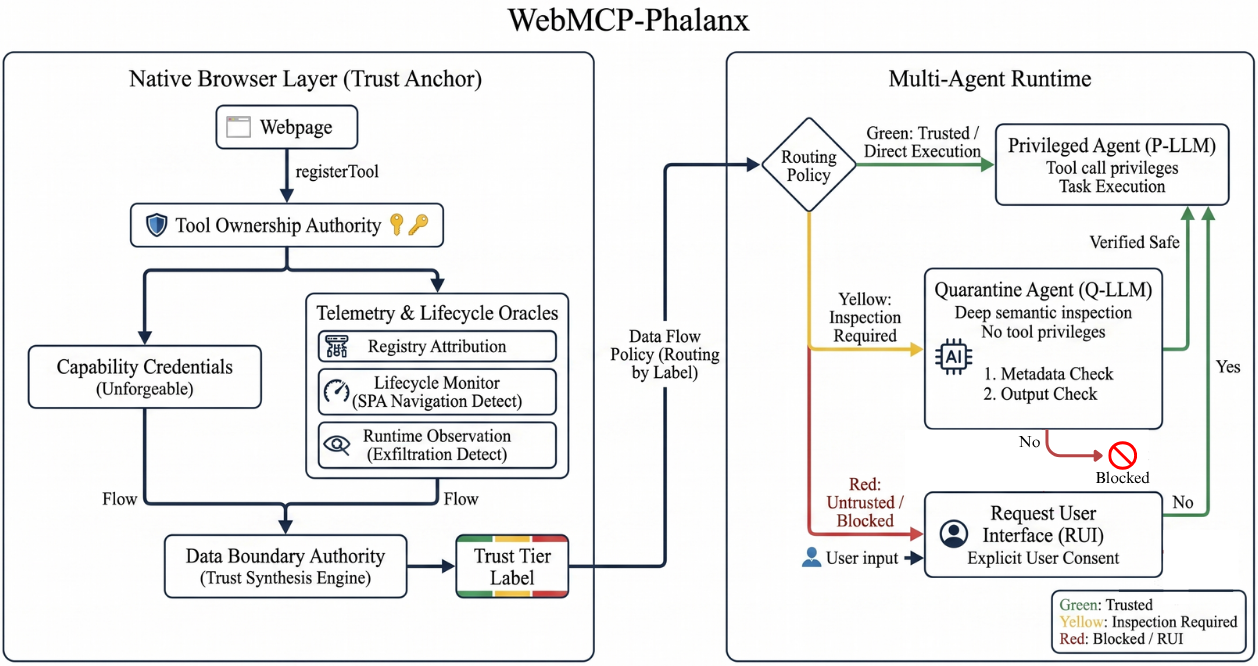} 
    \caption{The native browser side trust pipeline. The Tool Ownership Authority acts as the root of trust by issuing unforgeable capability handles, effectively isolating the registry from Turing complete same origin adversaries. The synthesis of telemetry and lifecycle states generates a deterministic Trust Tier Label. Note that while this architecture ensures ownership integrity, first registered name allocation remains a structural limitation (\S Limitations).}
    \label{fig:browser_pipeline}
\end{figure*}

The execution pipeline is divided into two continuous stages. As illustrated in Figure \ref{fig:browser_pipeline}, when a webpage script invokes \texttt{registerTool}, the underlying Tool Ownership Authority intercepts the call to physically manage state. The tool is then tagged with its origin and continuously evaluated by Telemetry and Lifecycle Oracles. A Data Boundary Authority synthesizes these streams into an immutable trust tier label. Finally, as shown in Figure \ref{fig:browser_pipeline}, the tool and its label are dispatched to the Multi Agent Runtime, which dynamically routes the payload to strictly enforce the assigned trust policy.

\subsection{Native Cryptographic Ownership Authority}
This module constitutes the root of trust for the entire framework. We completely abstract the core state of tool ownership into the browser's native implementation layer, uniformly governing tool registration, enumeration, and unregistration. This structural isolation renders the registry entirely inaccessible to Turing complete third party JavaScript.

Upon a successful registration request, the native execution engine mints an opaque, unforgeable capability handle. To subsequently revoke or overwrite a tool, the calling script must present this exact cryptographic credential. Consequently, a script—including a same origin third party adversary—can only mutate tools it natively registered. This capability based model establishes absolute ownership integrity rather than subjective attribution: it cryptographically proves continuous possession, not identity.

\subsection{Browser Side Telemetry and Lifecycle Oracles}
Operating directly atop the ownership layer, these native sub components continuously monitor execution state to feed deterministic signals into the Data Boundary Authority.

\begin{itemize}
\item Registry Attribution: Upon a tools/list invocation, the native browser retrieves the tool's registrant origin from the isolated core state and appends a spoof proof origin tag. While this definitively identifies the origin, it structurally cannot distinguish between discrete scripts sharing that origin (see Discussion).

\item Lifecycle Monitor: At registration, tools explicitly declare their lifecycle scope as either session or navigation. By natively intercepting the History API (pushState, replaceState, popstate), this module accurately detects logical page transitions in Single Page Applications (SPAs) and aggressively purges navigation scoped tools. This mitigates a critical specification gap where tools persist until a hard reload, forcefully aligning the tool's lifecycle with the page as perceived by the user.

\item Runtime Observation: During active tool execution, this module asynchronously monitors outbound exfiltration vectors (fetch, sendBeacon, Image.src) and DOM mutations via an isolated MutationObserver. These traces are compared against the tool's statically declared behavior. Upon detecting divergence (e.g., a self reported read only tool initiating a network fetch), the system immediately downgrades the trust label to Red. This recomputed label is injected into the execution output and future registry reads, guaranteeing the agent perceives the violation. This mechanism provides robust detect and contain semantics without requiring heavyweight sandbox isolation.
\end{itemize}

\subsection{Data Boundary Authority (Trust Synthesis Engine)}
This module deterministically synthesizes the telemetry signals into a singular Trust Tier Label for every tool. The algorithmic determination integrates two distinct dimensions:
\begin{itemize}[leftmargin=*]
    \item Origin Dimension (Derived from Ownership \& Attribution): Evaluates whether the tool is anchored by a valid capability credential issued by the native layer.
    \item Content Dimension (Derived from Telemetry \& Declarations): Evaluates the declared behavioral intent against the physical execution traces observed by the runtime.
\end{itemize}

The framework enforces three strict trust tiers:
\begin{description}[leftmargin=*]
    \item[Green (Trusted):] The tool possesses a cryptographically verified origin, and its runtime execution strictly adheres to its declared behavior. Policy: Permitted for direct execution.
    \item[Yellow (Requires Attention):] The tool possesses a verified origin, but its declared schema involves mutating its local environment. Policy: Subject to lightweight semantic verification.
    \item[Red (Untrusted):] The tool lacks a legitimate origin, or runtime telemetry has captured behavioral violations. Policy: Quarantined or explicitly blocked.
\end{description}

Crucially, the Origin Dimension is designed to exert absolute veto power over self reported attributes. However, as demonstrated in our evaluation, this veto mechanism cannot trigger against a same origin adversary, because every successfully registered tool inherently secures a valid capability.

\subsection{Asymmetric Dual Agent Isolation Runtime}
The downstream execution environment operates on an asymmetric multi agent topology comprising two distinct roles: the Privileged Agent (P), which holds explicit tool calling permissions, and the Quarantine Agent (Q), which is strictly isolated from execution privileges. 

Trust labels dictate the data flow topology: Green labeled content is routed directly to P for frictionless task execution. Conversely, untrusted or suspicious content is forcefully intercepted and routed into Q's isolated sandbox for semantic evaluation. Consequently, adversarial instructions concealed within suspicious payloads are absorbed entirely by Q, physically preventing the manipulation of P's control flow.

To counter semantic injections that bypass metadata labeling, Q enforces deep content inspection at two critical execution checkpoints:
\begin{itemize}
\item Checkpoint 1 (Pre Execution Description Analysis): Q evaluates the tool's static description and input schema during registration, successfully intercepting static prompt injections designed to manipulate agent planning.
\item Checkpoint 2 (Post Execution Return Analysis): Because a structurally benign tool may dynamically generate a malicious payload during execution, Q intercepts and sanitizes the returned context before it is permitted to re enter P's operational memory.
\end{itemize}

\subsection{Request User Interface}
The Request User Interface (RUI) on the left side of the architecture diagram serves as the final line of defense within the framework. Under extremely sensitive operations or in scenarios defying deterministic classification, the browser issues a secure confirmation request directly to the user via a trusted path. 

Consent is only triggered at the Red level; Green/Yellow levels are handled autonomously by the system's dual agent runtime, avoiding constant interruptions and preventing user fatigue. Consent is thus structurally preserved as the ``final line of defense'' rather than the primary criterion for security judgment.

In summary, the design of WebMCP-Phalanx is guided throughout by a single principle: the separation of trusted origin and trusted usage. The browser side establishes tool labels using an unforgeable, deterministic mechanism, while the multi agent side and data flow policies make semantic layer judgments strictly based on these labels. The former resolves ownership integrity and lifecycle attacks, while the latter targets semantic level injection attacks.
\section{Experiments}
\subsection{Experimental Setup}

All experiments were conducted in a real world browser environment, based on the reference implementation of WebMCP (@mcp-b/global 2.3.2, applying the 2026-06-17 draft specification). Attacks were injected via scripts to simulate a third party script attacker. An attack counts as successful when the agent calls the malicious tool and the receiving end receives the exfiltrated data. All agents in the multi agent module use the same language model, identical to the one used in the injection attack baseline we adopt, so that any performance variance stems from the architecture and labeling rather than a more powerful model. We deliberately refrain from blocking detected violations, in order to obtain an upper bound on attack success rate; this is a methodological choice, since the browser side labels and propagates by design while enforcement resides on the agent side.

\subsection{Label Driven Injection Defense }

We design a four layer ablation to evaluate how the multi agent system uses trust labels to defend against semantic injections the labels themselves cannot detect: L0 (No Defense), L1 (Labels Only), L2 (Rule based Stripping), and L3 (Multi Agent Content Judgment).For tools without a legitimate origin, the labels suffice; for legitimate tools carrying malicious content, Table 1 reports the attack success rates across the four layers. L1 confirms the labels' blind spot to content injection: with labels alone, 89\% of description injections and 48\% of return injections still penetrate, close to the undefended baseline. L2 reduces description injection to 25/80 through content agnostic redaction, and L3 closes that channel entirely (0/80); return injection is reduced to a 2/80 residual by both.Task completion (distinct from the attack success counts in Table 1) declines from 66/80 (L0) to 56/80 (L3) on the description channel, but this decline is not a defense cost: a no attack baseline completes 61/80, statistically indistinguishable from L3 (p = 0.48), while the higher L0 value stems from the checkout scenario, where the malicious tool's task fitting name scaffolds the agent into a fuller workflow. The scaffold survives description redaction (L2) and is thus carried by the tool name alone, so we anchor all completion comparisons to the no attack baseline.

\subsection{Injection Vector: Description vs. Returned Content}

Semantic injection can hide in the description at registration or in the returned content after execution, the latter being dynamic and absent at registration time. We added returned content injection and a corresponding post execution handling point. In Table 1, L1 is close to L0 for both vectors, because the labels are minted at registration while the malicious content of return injection does not yet exist; L1 for description injection is slightly higher than L0, since the labels mark the tools as trusted and thereby feed the agent misleading information. Returned content can be entirely replaced with placeholders, so rule based stripping prevents it almost completely; but after stripping, the tool name remains, and if it fits the task the agent still calls it, so only multi agent content evaluation closes the description channel entirely (Table 4).

\begin{table}[t]
\centering
\small
\begin{tabular}{@{}lcccc@{}}
\hline
Injection Vector & L0 & L1 & L2 & L3 \\
\hline
Description (Registration) & 69/80 & 71/80 & 25/80 & \textbf{0/80} \\
Returned Content (Execution) & 42/80 & 38/80 & 2/80 & \textbf{2/80} \\
\hline
\end{tabular}
\caption{Attack success counts for description and returned content injection across the four defense levels (n=80 per cell).}
\label{tab:vector}
\end{table}

\begin{table*}[t]
\centering
\small
\begin{tabular}{l|l|p{0.55\textwidth}}
    \textbf{Handling Point} &
    \textbf{Mechanism} &
    \textbf{Fails When} \\ \hline
    \makecell[l]{Registration:\\Description} & Label Rules &
    Always, under a same origin adversary: every successful registrant obtains a capability, so the origin dimension is constant and its veto cannot fire (Table in Appendix). \\ \hline
    \makecell[l]{Registration:\\Description} & \makecell[l]{Rule based\\Stripping} &
    The tool name is task relevant. ASR rises from 0\% to 70\% as name relevance increases (Table~\ref{table5}). \\ \hline
    \makecell[l]{Registration:\\Description} & \makecell[l]{Multi Agent\\(Q)} &
    Not observed; 0\% across all name relevance tiers (Table~\ref{table5}). \\ \hline
    \makecell[l]{Execution:\\Return Content} & Label Rules &
    Always: the label is minted at registration, whereas the injected content does not yet exist at that moment. \\ \hline
    \makecell[l]{Execution:\\Return Content} & \makecell[l]{Rule based\\Stripping} &
    The agent invokes the tool on its name before the carrier executes (1 of the 2/80 residual; the other is channel consistent). \\ \hline
    \makecell[l]{Execution:\\Return Content} & \makecell[l]{Multi Agent\\(Q)} &
    The agent invokes the tool on its name before the carrier executes (2/80).
\end{tabular}
\caption{Failure conditions of each mechanism at each handling point.}
\label{table2}
\end{table*}
Table 2 breaks down the failure condition of each mechanism. The 2/80 residuals at L2 and L3 are disjoint, and the matching rates are coincidental. Three of the four runs are name driven: the agent invoked vendor sync in the same turn as, or before, the carrier executed, so no returned content existed to strip or read (Q detects the carrier 40/40). The other two are not detection failures: the exfiltration tool presents a clean description and returns a benign status object, making it structurally invisible to both content checkpoints. What the content check intercepts is the carrier that induces the call, not the induced call itself; closing this residual requires a name layer or call timing gate.

\subsection{Cross Page Semantic Residual}
Lifecycle monitoring clears residual tools after a page transition, but the content those tools previously generated has already entered the agent's conversation history and continues to influence its later behavior. In the attack scenario, a tool on Page A injects instructions within its returned content to direct a bank transfer; after the agent navigates to Page B the tool is cleared, and we observe whether the residual semantics still drive the transfer.
A key finding from Table 3 is that merely clearing the tool only reduces the attack success rate from 80\% to 70\%, because the injected semantics reside in the conversation history, independent of whether the generating tool still exists, a pure semantic residual. Clearing the tool also lowers completion from 19/20 to 16/20, but not because clearing harms the task: in all such runs the agent believes the residual audit claim, halts, and asks the user to confirm a fabricated transfer instead of sending the payment. Our countermeasure extends the lifecycle of trust labels to the content the tools generate, invalidating the labels of any residual returned content in the agent's history when the tool is cleared; this reduces the attack success rate to 0\% without harming normal navigation, and simultaneously improves utility over clearing alone, since an attack succeeding during a page transition often distracts the agent from the original task, and removing the interference lets it stay focused and complete.

\begin{table}[t]
\centering
\small
\begin{tabular}{l|c|c}
    \textbf{Condition} & 
    \makecell{\textbf{Attack Success}\\\textbf{Rate}} & 
    \makecell{\textbf{Task Completion}\\\textbf{Rate}} \\ \hline
    No Defense & 80\% & 95\% \\ \hline
    No Attack & 0\% & 100\% \\ \hline
    \makecell[l]{Lifecycle Monitor Only} & 70\% & 80\% \\ \hline
    \makecell[l]{Lifecycle Monitor +\\Our Framework} & 0\% & 100\%
\end{tabular}
\caption{Cross Page Semantic Residual Analysis.}
\label{table4}
\end{table}

\subsection{Name Engineering \& Architectural Trade offs}

Table 4 shows that the limitation of rule based stripping against description injection stems from the agent calling a tool prematurely when its name fits the task, before content filtering intervenes. By fixing the injected content and varying only the name relevance, we quantify its impact on L2 and L3. The contrast is clear: the more closely the name fits the task, the weaker rule based stripping becomes, whereas the multi agent system reads the actual content and is unaffected by how the name is packaged. This effect is further modulated by substitutability: the checkout scenario is the most hijacked (20/20 at L0) because its malicious tools share the checkout's function and displace it by substitution, while the driver installation scenario, whose check–download–execute sequence offers no such substitute, is the only one where attacks frequently fail without any defense (10/20 at L0, all failed runs still completing the task). Substitutability, not sequence rigidity alone, governs reach.

\begin{table}[t]
\centering
\small
\begin{tabular}{l|c|c}
    \textbf{Name Relevance} & 
    \textbf{L2 ASR} & 
    \textbf{L3 ASR} \\ \hline
    N0 Completely Irrelevant & 0\% & 0\% \\ \hline
    N1 Generic & 0\% & 0\% \\ \hline
    N2 Common Step & 25\% & 0\% \\ \hline
    N3 Task Domain & 55\% & 0\% \\ \hline
    N4 Essential Step & 70\% & 0\%
\end{tabular}
\caption{Name Relevance on Defense Performance.}
\label{table5}
\end{table}
\section{Discussion}

The architectural foundation of WebMCP-Phalanx rests on the strict decoupling of provenance verification from semantic execution: the native browser layer establishes an unforgeable trust label, while the multi agent runtime enforces semantic layer policies based strictly on that anchor. This addresses a fatal structural gap in contemporary multi agent defenses. Frameworks such as Dual-LLM and CaMeL operate under the assumption that the system inherently knows which content is untrusted, yet they fundamentally fail to resolve where that label originates or whether it can be cryptographically forged by an adversary. In a multi party web environment governed by a flat Same Origin Policy, this trusted oracle does not naturally exist. Phalanx physically constructs this oracle at the native browser layer, and our empirical measurements formally delimit the exact reach of a capability based anchor.

\subsection{The Cryptographic Reach of Bearer Capabilities.}
Ownership integrity is empirically validated and enforced: a third party script holding a valid capability handle is strictly isolated to modifying its own tools. Attempts to manipulate legitimate tools (e.g., checkout) deterministically raise an \texttt{InvalidStateError}. Consequently, overwrite and revocation attacks (V4 and V6) drop from 20/20 to 0/20, while V5's legitimate response rate rises from 0/20 to 20/20. 

However, we formally demonstrate that subject attribution is structurally beyond the reach of a bearer capability. This is not an artifact of our implementation; the native layer would issue a handle to a malicious \texttt{partner.js} on exactly the same terms. This fundamental gap cannot be repaired at the JavaScript execution layer and must be structurally addressed within the WebMCP specification through two explicit mechanisms. First, Native Script Provenance: JavaScript lacks unforgeable script identity. Because properties like \texttt{document.currentScript} and stack introspection are trivially spoofable by a same origin adversary, page layer attribution is inherently insecure and must be supplied by the browser's native compilation pipeline. Second, a Declarative Trust Topology: A third party SDK is loaded by the site's own choice, yet the browser possesses no observable heuristic to distinguish ``loaded by'' from ``trusted by.'' This operational intent exists exclusively within the site's logic and must be explicitly declared, analogous to a Content Security Policy (CSP). We strongly recommend the WebMCP specification incorporate both primitives; evaluating their native implementations falls outside the scope of this paper.

\subsection{Adversarial Confounders in Utility Measurement.}
The critical value of cryptographically binding content trust to the origin's lifecycle is most evident in cross page evaluations: once a tool is cleared by the SPA router, its previously generated payload still lingers in the agent's context window. Only our strict lifecycle binding mechanism systematically eradicates this semantic residual. 

Furthermore, our cross page and completion analyses expose a critical statistical confounder in agentic evaluation: utility under an adversary is not a neutral metric. An active attack can artificially depress task utility by distracting the agent, or paradoxically inflate it because the malicious tool's task fitting nomenclature acts as an adversarial scaffold, successfully guiding the agent through a more complete workflow ($p=0.003$). Consequently, measuring task completion against an attacked baseline risks fundamentally misattributing either effect to the defense mechanism itself. To ensure statistical rigor, we mandate anchoring all utility comparisons exclusively to a zero attack baseline.

Ultimately, the two lines of defense in Phalanx reflect the fundamental dichotomy of the WebMCP threat landscape. Specification level protocol flaws demand deterministic, system level eradication independent of the underlying LLM, whereas semantic prompt injections necessitate the deep contextual comprehension provided by an isolated, multi agent evaluation runtime.
\section{Conclusion and Future Work}

Our reference implementation operates as a JavaScript polyfill, so two
security assumptions are emulated rather than natively implemented. The
first is API surface integrity, which we emulate by preemptively loading
and caching pristine built-in references. A native layer would secure the
API surface before any page script executes, which requires the
specification to enforce API immutability and tamper-proof native
bindings. Because the polyfill operates under a strictly weaker privilege
condition than a native implementation, our defense metrics establish a
conservative lower bound. The second is cryptographic script provenance,
which is structurally absent in vanilla JavaScript. Results involving
provenance therefore rest on an architectural existence assumption rather
than a cryptographically transferable bound. A capability also acts as a
bearer credential and remains vulnerable to the confused deputy problem:
a registrant that leaks its handle, or that exposes a function revoking
tools on its behalf, forfeits the ownership guarantee.

Two further components are presented as architectural primitives for the
specification and are not empirically evaluated: the site-declared trust
policy and the out-of-band Request User Interface. Runtime Observation is
a post-execution detection mechanism, so the first-order side effects of
an initial violation cannot be prevented. Within the observation window
of an asynchronous \texttt{execute} operation, the monitor may also fail
to attribute DOM mutations to a specific tool, leading to attribution
collisions.

WebMCP remains in its formative drafting stages. We hope the
vulnerabilities and architectural remedies reported here will assist the
W3C in embedding security-by-design principles before widespread
deployment. Two directions remain open. First, resilience against adaptive
adversaries: our primary evaluation measures fixed, known injection
techniques, and we additionally modeled a white-box adaptive attacker
endowed with the Quarantine Agent's classifier prompt and with end-to-end
success feedback. This attacker bypassed 15 of 16 description-layer
payloads with a median of one rewrite round, succeeding not by defeating
Q's semantic judgment but by relocating the payload onto the task-fitting
tool name, an adversarial scaffold that lies outside the bounds of
content inspection. Closing this vector requires a name-layer or
call-timing gate. Second, deterministic execution sandboxing: Runtime
Observation is optimized for detection, containment, and disclosure
rather than prevention, and blocking covert behavior during internal tool
execution demands sandbox-level isolation at the browser layer.

\bibliographystyle{unsrtnat}
\bibliography{references}

\end{document}